\documentclass[notitlepage]{revtex4-1}
\usepackage{amsmath}
\usepackage{graphicx}
\usepackage{subcaption}
\usepackage{hyperref}
\hypersetup{
    colorlinks=true,        
    linkcolor=blue,         
    citecolor=blue,         
    urlcolor=blue           
}

\usepackage{bm}
\usepackage{algorithm}
\usepackage{algpseudocode}
\usepackage[english]{babel}
\usepackage[autostyle, english = american]{csquotes}
\MakeOuterQuote{"}

\begin{document}

\title{Equilibrium-like statistical mechanics in space-time for a deterministic traffic model far from equilibrium} 

\author{Aryaman Jha$^{1}$, Kurt Wiesenfeld$^{1}$, Jorge Laval$^{2}$}

\affiliation{$^{1}$Center for Nonlinear Sciences, School of Physics, Georgia Institute of Technology}
\affiliation{$^{2}$Traffic Flow Laboratory, School of Civil and Environmental Engineering, Georgia Institute of Technology}

\date{\today}

\begin{abstract}
    Motivated by earlier numerical evidence for a percolation-like transition in space--time jamming, we present an analytic description of the transient dynamics of the deterministic traffic model elementary cellular automaton rule~184 (ECA184). By exploiting the deterministic structure of the dynamics, we reformulate the problem in terms of a height function constructed directly from the initial condition, and obtain an equilibrium statistical mechanics-like description over the lattice configurations. This formulation allows macroscopic observables in space--time, such as the total jam delay and jam relaxation time, as well as microscopic jam statistics, to be expressed in terms of geometric properties of the height function.  We thereby derive the associated scaling forms and recover the critical exponents previously observed in numerical studies. We discuss the physical implications of this space--time geometric approach.
\end{abstract}

\maketitle
\section{Introduction}

Equilibrium statistical mechanics provides a general framework for relating microscopic dynamics to macroscopic behavior, including the fluctuation properties \cite{kardar2007statistical, kardar2007statistical2,huang2008statistical}. For systems out of equilibrium, no comparably general framework exists, and basic organizing principles often depend on the class of dynamics under consideration \cite{derrida2007non, marconi2008fluctuation, bertini2015macroscopic}. There is nevertheless broad interest in equilibrium-like descriptions for nonequilibrium systems with approaches such as macroscopic fluctuation theory \cite{bertini2002macroscopic} and stochastic thermodynamics \cite{jarzynski1997nonequilibrium,seifert2005entropy, seifert2012stochastic}. There have also been efforts to obtain thermodynamic descriptions for non-thermal systems \cite{reiss1986thermodynamic}. 

Cellular automaton models of vehicular traffic provide a class of nonequilibrium systems with active (self-driven) motion \cite{nagel1992cellular, biham1992self, chowdhury2000statistical}. A standard example is the one-dimensional Nagel--Schreckenberg (NS) model, where vehicles accelerate up to a maximum velocity $v_{\max}$ and undergo stochastic braking with probability $p$ \cite{nagel1992cellular}. In the special case $v_{\max}=1$, the NS model reduces to the totally asymmetric simple exclusion process (TASEP), in which particles hop unidirectionally subject to hard-core exclusion \cite{derrida1992exact, derrida1993exact}. The deterministic limit of TASEP corresponds, in Wolfram's taxonomy,  to elementary cellular automaton rule 184 (ECA184) \cite{wolfram2003new}.

In the statistical-physics literature, traffic models are most often analyzed in terms of densities, currents, and fluctuations of hydrodynamic variables \cite{helbing2001traffic, schadschneider2010stochastic}. By contrast, traffic-flow research emphasizes a geometric description based on space--time plots of vehicle trajectories, focusing on the spatial extent, temporal duration, and accumulated delay of congested regions \cite{edie1963discussion, gartner2002traffic, laval2023self}. This geometric perspective was explored by Nagel and Paczuski in the Nagel--Schreckenberg model through the fractal structure of jams \cite{nagel1995emergent}. More recently, motivated by physical quantities of interest to traffic flow community we applied a similar viewpoint to the simpler deterministic system ECA184 in its transient regime by treating jammed regions in space--time as connected clusters \cite{jha2025simple}. We found that the resulting cluster statistics exhibit scaling consistent with a percolation transition, with critical exponent values indicative of a comparatively simple underlying theory.

In this manuscript, we develop an analytic statistical description of ECA184. By exploiting specific properties of the deterministic dynamics, we obtain a ``micro-canonnical ensemble" over different initial lattice configurations in terms of an energy-like functional analogous to equilibrium systems, and derive scaling forms which recover the computationally observed critical exponents. We also speculate how the space--time viewpoint may be connected to an equilibrium thermodynamic picture.

This paper is organized as follows. 
 In Sec.~\ref{sec:bgg} we summarize the relevant phenomenology reported in our previous numerical study and introduce a new, key quantity we call the height function $H(X)$. In Sec.~\ref{sec:statform} we develop an analytic statistical description in the large system size limit ($L\rightarrow\infty$) and derive the corresponding probability measures. In Sec.~\ref{sec:scalexp} we obtain the scaling forms and use these to derive the critical exponent values. Finally, in Sec.~\ref{sec:summart_disc} we reinterpret our results and discuss further physical and formal implications. In Appendix~\ref{app:appA} we numerically check key analytical assertions and we defer some details of the scaling calculations to Appendix~\ref{app:scalingDerivation}.


\section{Background}\label{sec:bgg}

\subsection{Setup and Phenomenology}\label{subsec:setup_phenom}

ECA184 consists of a one-dimensional lattice of length $L$ with periodic boundary conditions, representing a single-lane road. Each site is either occupied by a vehicle ($1$) or empty ($0$). Vehicles are initialized randomly with a fixed density $\rho$, corresponding to exactly $\rho L$ occupied sites. Time evolves in discrete steps according to a deterministic update rule: at each step, a vehicle moves forward by one lattice site if the site ahead is empty; otherwise, it remains stationary. The dynamics conserve the total number of vehicles, so $\rho$ acts as a control parameter.

We focus on the transient dynamics generated by random initial conditions, and in particular on the geometric structures that appear in space--time before the system reaches its steady state by at most time $T_{\text{max}} = L/2$. By analyzing these transient space--time patterns, we find that the system exhibits scaling behavior consistent with a percolation-type transition as the vehicle density is varied \cite{jha2025simple}.

The evolution of the system can be visualized using a space--time diagram, where each horizontal row represents the lattice configuration at a given time with the tuple $(X,T)$ representing a lattice site $X$ at time $T$. In these diagrams, jammed regions appear as connected clusters of occupied sites extending over both space and time. These clusters represent traffic jams: vehicles inside a cluster are temporarily unable to move freely, and the cluster persists until it is eliminated by interactions with surrounding empty regions. Note that the maximum total space--time area until the maximum possible relaxation time is given by $T_{\text{max}}L = \frac{L^2}{2}$

These connected clusters map to physical observables in traffic, in particular: the area $a_i$ represents the total time spent stationary by cars in the $i^{th}$ cluster; the cluster height $\theta_i$ represents the lifetime of that jam (see Fig~\ref{fig:example_microscopic}) In addition to jam clusters, we similarly define \emph{void clusters} as connected regions of empty sites in space--time, with corresponding lifetimes $\bar{\theta}_i$. 

These clusters represent \textit{microscopic} features; we also define (global) \textit{macroscopic} quantities, {\it i.e.} the total delay over all clusters $A = \sum_{i} a_i$ and the time it takes the whole system to relax $T_R = \max_{i} \theta_i$. Further the normalized delay $\phi = A/(\frac{L^2}{2})$ acts as an order parameter and $T_R$ is analogous to a system response when the behavior is averaged over an ensemble of initial conditions.

A useful microscopic description of the space--time geometry is obtained by decomposing each jam cluster into simpler diagonal components, which we refer to as \emph{elementary jams}. Each elementary jam has a well-defined length $m_i$. As illustrated in the sample realization shown in Fig.~\ref{fig:example_microscopic}, the total number of elementary jams is equal to the number of vehicles -- that is, each initially occupied site anchors a unique elementary jam in the space-time plot.  All macroscopic observables and microscopic cluster properties can be expressed directly in terms of the collection of elementary jams associated with a given initial condition, for example: 

\begin{figure}[h]
    \centering
    \begin{subfigure}[t]{0.65\columnwidth}
        \centering
        \includegraphics[width=\linewidth]{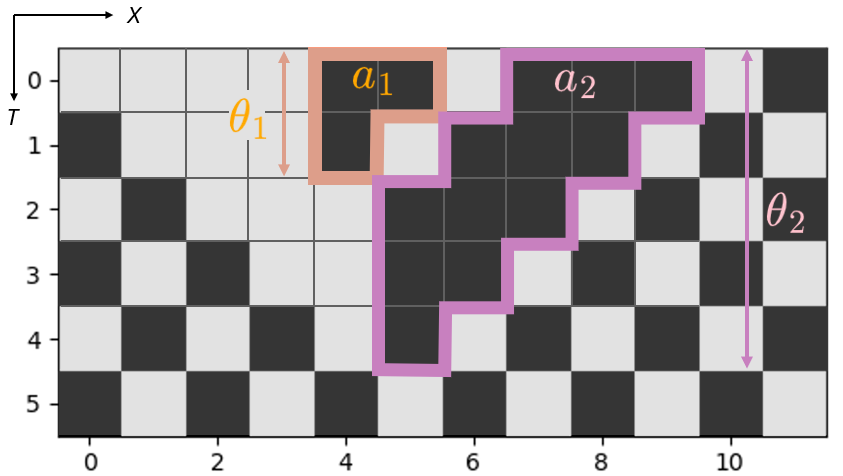}
        
    \end{subfigure}
    \hfill
    \begin{subfigure}[t]{0.3\columnwidth}
        \centering
        \includegraphics[width=\linewidth]{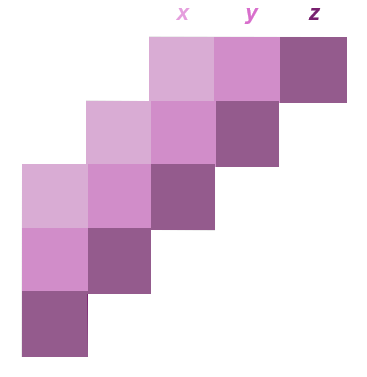}
    \end{subfigure}
    \caption{(Left) Spacetime plot of a system of size $L = 12$ and filling density $\rho = 0.5$, showing two jam clusters. The cluster areas $a_i$ and lifetimes $\theta_i$ are indicated by the arrows. 
 (Right) Decomposition of the largest cluster into elementary jams of lengths $m_x, m_y$ and $m_z$.}
    \label{fig:example_microscopic}
\end{figure}

\begin{equation}\label{eqn:elem}
A = \sum_j m_j , \quad \text{and}\quad
T_R
=
\begin{cases}
\max_j m_j, & \rho \leq \tfrac{1}{2}, \\
\max_j \bar{m}_j, & \rho > \tfrac{1}{2},
\end{cases}\:, 
\end{equation}
where for densities above $\rho_c = 1/2$, void clusters play the dominant role in controlling the relaxation dynamics.

In Ref.~\cite{jha2025simple}, we studied the dependence of these observables on the density $\rho$ by sampling uniformly over all $\binom{L}{\rho L}$ initial conditions with fixed $L$ and $\rho$. For a generic observable $\mathcal{O}$ defined on the space--time evolution, ensemble averages are computed as
\begin{equation}\label{eqn:space_time_avg}
\langle \mathcal{O} \rangle
= \sum \mathcal{O}[X,T]\; P[X,T],
\end{equation}
where $\cdot[X,T]$ denotes the functional over the entire space--time plot, $P[X,T]$ denotes the probability of a given space--time configuration and the sum $\sum$ runs over realizations generated from the ensemble of initial conditions. 

Finite size scaling analysis of the numerical data suggests that, in the ``thermodynamic limit" ($L\rightarrow\infty$), the order parameter $\phi$ vanishes above criticality while the relaxation time diverges on approach to $\rho_c$ from either side as (in terms of $\Delta = \rho-\rho_c$)
\begin{equation}
\phi \sim \Delta^{\beta}, \quad \beta = 1.0(0), \qquad\text{and}\qquad
\langle T_R\rangle \sim |\Delta|^{-\gamma}, \quad \gamma = 2.0(4)
\end{equation}
where of $\Delta = \rho-\rho_c$ , and the parenthesis signify the numerical uncertainty in the trailing decimal place. Since $T_R$ is the lifetime of the longest-lived jam cluster for $\rho\le \rho_c$ or void cluster for $\rho>\rho_c$, $\langle T_R\rangle$ plays the role of a temporal correlation length. Consistent with this interpretation, the divergence exponent satisfies
\begin{equation}
\xi \sim \Delta^{-\nu} \qquad\text{and}\qquad\nu = 2.0(2),
\end{equation}
where $\nu$ is the correlation-length exponent.

At the microscopic level, the distributions of cluster lifetimes $\theta$, cluster areas $a$, and elementary jam lengths $m$ follow power laws at criticality ($\Delta = 0$),
\begin{equation}
p'(y,\Delta = 0) \sim y^{-\tau}, \qquad y \in \{\theta, a, m\}, \quad \tau = 1.5(0).
\end{equation}
Away from criticality, these distributions acquire a density-dependent cutoff and take the scaling form
\begin{equation}
p'(y,\Delta) \sim y^{-\tau}\,F\!\left(y\,|\Delta|^{1/\sigma}\right),
\qquad \sigma = 0.5(3).
\end{equation}
The exponents are not independent. In particular, they satisfy the relations (within numerical certainty)
\begin{equation}
\gamma = \nu = \frac{1}{\sigma}\qquad \text{and}\qquad\beta = \frac{\tau-1}{\sigma} = \nu(\tau-1)
\end{equation}

\begin{figure}[h]
    \centering
    \begin{subfigure}[t]{0.45\columnwidth}
        \centering
        \includegraphics[width=\linewidth]{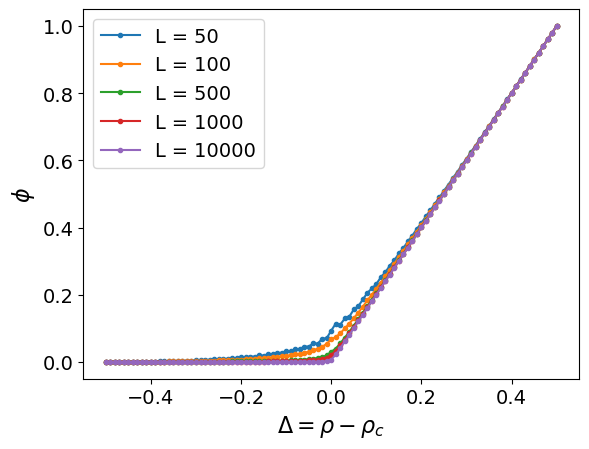}
        \caption{}
        \label{fig:example_macroscopic_a}
    \end{subfigure}
    \hfill
    \begin{subfigure}[t]{0.47\columnwidth}
        \centering
        \includegraphics[width=\linewidth]{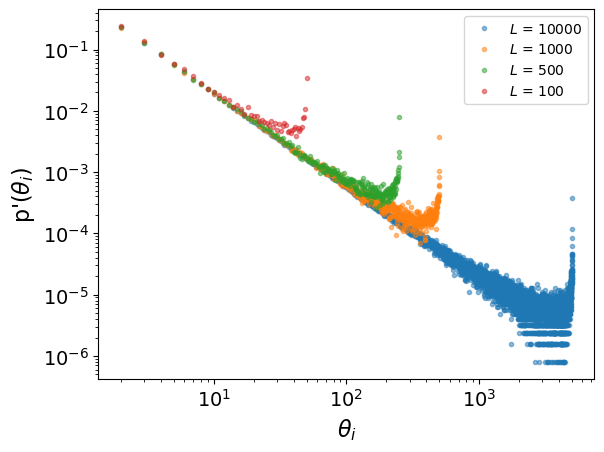}
        \caption{}
        \label{fig:example_microscopic_b}
    \end{subfigure}
    \caption{Observables: (a) (macroscopic) order parameter $\phi$ for different system sizes $L$ (b) (microscopic) cluster lifetime distribution $p'(\theta)$ for different $L$}
    \label{fig:observable}
\end{figure}

Some example observables are shown in Fig.~\ref{fig:observable}. Taken together, these results show that transient jamming in ECA184 can be understood as a space--time clustering phenomenon with scaling properties characteristic of a percolation transition.


\subsection{Height function $H(X)$}

Since the dynamics is deterministic, any observable is a function of the initial condition. In principle it is possible to obtain the observables directly from the initial condition without obtaining solutions to the governing dynamics., and we may rewrite Eqn. \ref{eqn:space_time_avg} solely in terms of the initial condition,
\begin{equation}\label{space_only}
    \langle \mathcal{O}\rangle = \sum_{\{X\}} \mathcal{O}[X]P[X]
\end{equation}
where $P[X]$ is given by the uniform distribution $\frac{1}{\binom{L}{\rho L}}$ over the set of initial conditions. In practice the mapping between initial conditions and observables for most deterministic systems is not known. However, for our case we can obtain such a mapping through a function $H(X)$ called the height function, such that we can exactly calculate the observables as geometric properties of $H$.

We construct $H(X)$ as follows \cite{laval2023self, jha2025simple}. 
 We start with a related height construction common in the literature linking TASEP and KPZ \cite{johansson2000shape} which we denote as $H_{\text{KPZ}}(X,T)$. Starting from the initial configuration of length $L$, we first reverse the order of sites and (for convenience) interpret the sequence as steps of a walk by replacing each $1$ with a step $+1$ and each $0$ with a step $-1$. This gives $H_{\text{KPZ}}$ at time $T = 0$ as
\begin{equation}
H_{\mathrm{KPZ}}(0,  0)=0, \qquad
H_{\mathrm{KPZ}}(X',T = 0)=\sum_{X'=1}^{X}I(X'), \quad X\ge 1,
\end{equation}
where $I(X')=\pm1$ denotes the step associated with the site at position $X'$.

Next, we shift the sequence so that the global minimum of $H_{\mathrm{KPZ}}(X,T=0)$ occurs at the origin. Let $X_{\min}$ be the smallest index at which this minimum is attained. Using this shift we obtain the height function $H(X)$ as
\begin{equation}\label{eqn:h_defn}
H(0)=0, \qquad
H(X)=\sum_{X'=1}^{X}I(X'-X_{\min}), \quad X\ge 1,
\end{equation}
with indices understood modulo $L$.

The function $H(X)$ will serve as the basic object from which all observables are obtained. Fig.~\ref{fig:ejam_hx} shows an example of $H(X)$ constructed using above procedure from the flipped space--time plot (inverted x-axis). 

Note that our $H(X)$ differs from $H_{\mathrm{KPZ}}(X,T)$ in that, in the deterministic limit of TASEP, $H(X)$ encodes the entire transient dynamics for a given initial condition. In particular, elementary jams and associated observables can be obtained exactly from $H(X)$, a property that is not available in the stochastic $H_{\text{KPZ}}(X,T)$ which evolves with the lattice.  Also note that at $\rho=0.5$, the number of upward and downward steps in $H(X)$ are equal, so that $H(0)=H(L)=0$ and $H(X)\ge 0$ for all intermediate $X$. Such paths are known as Dyck paths in the combinatorics literature \cite{stanley2015catalan}.

\subsection{Elementary jams from $H(X)$}

By construction, from Eqn.~\ref{eqn:h_defn} the height function may be viewed as the position of a one--dimensional random walk, with the spatial index $X$ playing the role of time. The walk advances in unit steps, moving up for occupied sites and down for empty sites in the initial configuration. A key property of this walk is that the elementary jams generated by the initial condition can be obtained directly from simple geometric features of $H(X)$. This correspondence between the geometry of the height function and elementary jams was previously exploited to develop an efficient computational algorithm for extracting elementary jams directly from initial conditions \cite{jha2025simple}.

\begin{figure}
    \centering
    \includegraphics[width=0.7\linewidth]{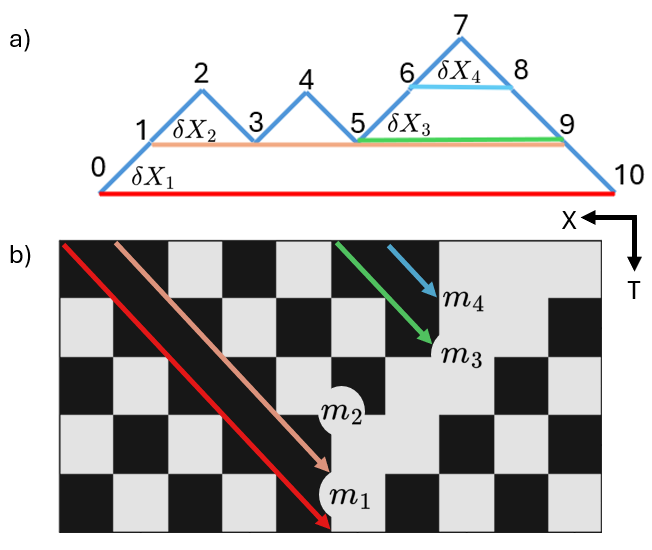}
    \caption{a) $H(X)$ for a system of size $L = 10$ for initial configuration $[0 0 0 1 1 0 1 0 1 1]$, the \textit{FCT} are shown visually as horizontal line segments with labels $\delta X_i$. b) The space-time plot (with X-axis flipped) for the given initial condition with the elementary jams with lengths $m_i$ colored similarly to the corresponding FCT $\delta X_i$. }
    \label{fig:ejam_hx}
\end{figure}

Specifically, elementary jams correspond to excursions of the walk above a given height. To formalize this, we define the \emph{first crossing time} (FCT):  for a chosen reference height, the first crossing time $\delta X_i$ is defined as the interval between the point where $H(X)$ first crosses above that height and the point where it subsequently crosses below it. In other words, $\delta X_i$ measures the horizontal extent of an excursion of $H(X)$ above the reference level.  (See  Fig.~\ref{fig:ejam_hx} for an example.)

Each such excursion corresponds to a single elementary jam. Owing to the unit slopes of jam and void boundaries in the space--time diagram, the length of the associated elementary jam is given by
\begin{equation}
m_i = \frac{\delta X_i}{2}.
\end{equation}

It is important to distinguish the first crossing time from a first return or first passage time. The walk may return to the same height multiple times during an excursion; however, an elementary jam persists until the walk crosses below the reference height for the first time. This distinction is illustrated in Fig.~\ref{fig:ejam_hx}: the trajectory starting at index~$1$ returns to the same height at indices~$3$, $5$, and~$9$, but crosses below that height for the first time only at index~$9$. Similarly, the excursion beginning at index~$5$ also terminates at index~$9$.

Importantly, since the elementary jams can be used to obtain all other quantities of interest, we may obtain all the observables directly in terms of $H(X)$ (Sec. \ref{subsec:observables_all}) .



\section{Statistical formulation in the thermodynamic limit $L\rightarrow\infty$}\label{sec:statform}
\subsection{Fokker--Planck description of $H(X)$}\label{subsec:fp_description}

We want to express macroscopic observables and microscopic distributions as functions of the control parameter $\Delta=\rho-\rho_c$. Since all observables can be expressed as geometric functionals of the height function, ensemble averages over initial conditions in Eqn.~\ref{space_only} may be rewritten as
\begin{equation}\label{hx_only}
\langle \mathcal{O}\rangle
= \sum_{[H(X)]} \mathcal{O}[H(X)]\,P[H(X)],
\end{equation}
where $[H(X)]$ denotes a functional dependence on the height profile.

Since $H(X)$ may be interpreted as the position of a one--dimensional random walk with lattice index $X$ playing the role of time (Eqn.~\ref{eqn:h_defn}), when initial conditions are sampled uniformly at fixed density $\rho$, this induces a stochastic process for $H(X)$. Let $P(H,X)$ denote the probability that the height takes value $H$ at position $X$. The process is Markovian and obeys
\begin{equation}\label{eqn:markov_finite}
P(H,X)=P_{\uparrow}\,P(H-1,X-1)+P_{\downarrow}\,P(H+1,X-1),
\qquad P(0,0)=1,
\end{equation}
where $P_{\uparrow}$ and $P_{\downarrow}$ are the probabilities of upward and downward steps.

In the limit $L\rightarrow\infty$, these probabilities are fixed by the density,
\begin{equation}
P_{\uparrow}=\rho, \qquad P_{\downarrow}=1-\rho,
\end{equation}
so that
\begin{equation}\label{eqn:markov_thermo}
P(H,X)=\rho\,P(H-1,X-1)+(1-\rho)\,P(H+1,X-1).
\end{equation}

In the same limit, we replace the discrete process by a continuum description \cite{ding2004first},
\begin{equation}
p(h,x)=\rho\,p(h-\mathrm{d}h,x-\mathrm{d}x)+(1-\rho)\,p(h+\mathrm{d}h,x-\mathrm{d}x),
\end{equation}
which upon Taylor expansion yields the drift--diffusion equation in $(h,x)$
\begin{equation}\label{eqn:fp_x}
\partial_x p(h,x)
=-2v_0\Delta\,\partial_h p(h,x)
+D_0\,\partial_{hh}p(h,x),
\end{equation}
with $\Delta = \rho - \frac{1}{2}$, $v_0=\mathrm{d}h/\mathrm{d}x$ and $D_0=(\mathrm{d}h)^2/(2\,\mathrm{d}x)$.
The solution to Eqn.~\ref{eqn:fp_x} is given by the broadening Gaussian,

\begin{equation} \label{eqn:fp_soln}
p(h,x)
=
\frac{1}{\sqrt{4\pi D_0 x}}
\exp\!\left\{
-\frac{\left(h - 2v_0\Delta x\right)^2}{4 D_0 x}
\right\}.
\end{equation}

The probability for the height $h(x)$ may be written in path-integral form as
\begin{equation}
p(h,x|h_0,x_0)
=\int_{h_0,x_0}^{h,x}\mathcal{D}h'\;p[h'(x')].
\end{equation}
where $\mathcal{D}h'$ denotes the path-integral measure and $p[\cdot]$ is the path functional over an entire $h'(x')$. We can express this as \cite{Wilkins_Rigopoulos_Masoero_2021}
\begin{equation}\label{eqn:path_int_formula}
p(h,x|h_0,x_0)
=\mathcal{N}\int_{h_0}^{h}\mathcal{D}h'\,
\exp\{-S[h'(x')]\},
\end{equation}
with action
\begin{align}
S[h'(x')]
&=\int_{x_0}^{x}dx'\Bigg\{
\frac{1}{4D_0}\big(\partial_{x'}h'(x')\big)^2
-\frac{1}{2D_0}\hat{U}\!\left(h'(x')\right)
+\partial_{x'}V(h'(x'))
\Bigg\},\\[4pt]
V(h)&=-2v_0\!\left(\rho-\tfrac{1}{2}\right)h,\\
\hat{U}(h)&=\frac{\gamma}{4}\partial_h^2V(h)
-\frac{1}{4}\big(\partial_h V(h)\big)^2,
\qquad \gamma=2D_0 .
\end{align}

Combining these expressions yields a quadratic action,
\begin{equation}\label{eqn:action_formula}
S[h(x)]
=\int_{x_0}^{x}dx'\left\{
K_0+K_1\,\partial_{x'}h(x')+K_2\big(\partial_{x'}h(x')\big)^2
\right\},
\end{equation}
with
\[
K_0=\frac{v_0^{2}}{D_0}\Delta^{2},\qquad
K_1=-2v_0\Delta,\qquad
K_2=\frac{1}{4D_0},
\qquad
\Delta=\rho-\tfrac{1}{2}.
\]

The probability of observing a given height function is therefore
\begin{equation}
\boxed{p[h(x)]=\frac{1}{Z}\exp\{-S[h(x)]\}},
\qquad
Z=\int\mathcal{D}h\;\exp\{-S[h(x)]\}.
\end{equation}

We pause here to consider a technical point.  Because the construction of $H(X)$ involves a cyclic shift, multiple initial conditions could map to the same height function. This introduces a degeneracy factor $w[h(x)]$ such that
\begin{equation}\label{eqn:weighted}
\langle \mathcal{O}\rangle_{\mathrm{ini}}
=\sum_{[h(x)]}\mathcal{O}[h(x)]\,p[h(x)]\,w[h(x)].
\end{equation}
We assert that $w[h(x)]=1$ in the thermodynamic limit, so that
\begin{equation}
\langle \mathcal{O}\rangle_{\mathrm{ini}}
=\sum_{[h(x)]}\mathcal{O}[h(x)]\,p[h(x)] .
\end{equation}

We return to this point in the discussion section.

\subsection{Observables from $H(X)$}\label{subsec:observables_all}

\subsubsection{Mircoscopic distributions: $\{ \theta, m\}$}\label{microscopic_distr_expo}

\begin{figure}
    \centering
    \includegraphics[width=0.7\linewidth]{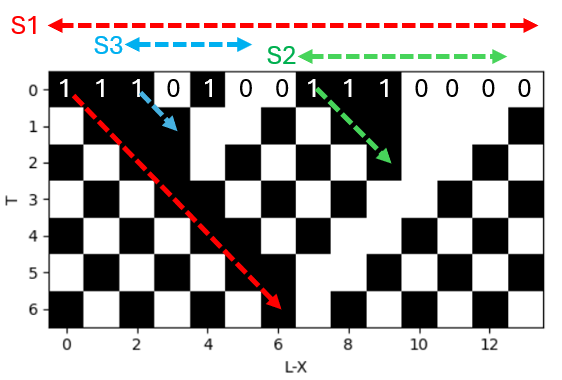}
    \caption{Example of a space–time plot showing three arbitrary subsequences $S1$, $S2$, and $S3$ of the initial condition. Each subsequence contains an equal number of 0s and 1s, and the colored diagonal arrows indicate the corresponding features on the space-time plot. $S1$, $S2$ and $S3$ corresponds to elementary jams, while $S1$ and $S2$ also correspond to jam lifetimes. }
    \label{fig:subseqreturn}
\end{figure}

We make the following observation: every elementary jam with a finite length terminates when it intersects an elementary void of the same length.  In the space-time plot, the elementary jam and its corresponding elementary void form the two equal sides of an isoceles triangle, whose third side is a substring of the initial condition (see Fig.~\ref{fig:subseqreturn}  for an illustration).  Thus, the length of this substring is twice the length of the elementary jam.  A similar observation may be made for the cluster lifetimes. These observations imply that the elementary jam lengths and the cluster lifetime distribution are given by the first return time distribution to Eqn.~\ref{eqn:markov_finite}.

For the example shown in Fig.~\ref{fig:subseqreturn} , we choose any arbitrary set of subsequences from the initial condition, say $S1$, $S2$, and $S3$; one can check that each of these has an equal number of 0s and 1s, and that the diagonal arrows show the corresponding elementary jams.  Furthermore, the lengths of the elementary jams corresponding to $S1$ and $S2$ are also cluster lifetimes (while $S3$ is not).

In the thermodynamic limit, we define $p'(\theta,\Delta)$ to denote the probability distribution for the jam lifetime $\theta$ at density $\rho$. In the continuum limit it satisfies the first return time problem for the stochastic process given by Eqn.~\ref{eqn:fp_x}. The first return time distribution $f(\delta x)$ is given by the well-known inverse Gaussian form \cite{redner2001guide,ding2004first}
\begin{align}
\label{eqn:fpt_solution}
f(\delta x) = \frac{h_0}{\sqrt{4\pi D_0 (\delta x)^3}}
\exp\!\left\{-\frac{\left(h_0-2v_0\Delta\; \delta x\right)^2}{4D_0\;\delta x}\right\},
\end{align}
where $h_0$ denotes the initial height of the jam.

Since the observable $\theta$ is related to the return time via $\theta = \delta x\;/2$, the jam lifetime distribution is
\begin{align}
\label{eqn:expression_microscopic}
p'(\theta,\Delta) = 2f(2\theta,\Delta)
&= \frac{h_0}{\sqrt{4\pi D_0 (2\theta)^3}}
\exp\!\left\{-\frac{\left(h_0-4v_0\Delta \theta\right)^2}{8D_0\theta}\right\}\\ &\approx \Big (\frac{h_0}{\sqrt{32\pi D_0}}\Big )\theta^{-3/2}
\exp\!\left\{-\frac{2v_0 ^2}{D_0}(\theta\Delta^2)\right\},\quad\theta>>1
\end{align}

$p'(\cdot)$ in Eqn.~\ref{eqn:expression_microscopic} also gives the elementary jam distribution by replacing $\theta$ with $m$.

Furthermore, from this we get the microscopic distribution for large $\theta,\;m$ at $\rho = 1/2$  ({\it i.e.} $\Delta=0$),

\begin{align}
  \boxed{ p'(y,\Delta=0)\sim y^{-3/2}} \qquad y = \{\theta,m\}
\end{align}

and for a general $\Delta$,

\begin{equation}
\boxed{p'(y,\Delta) \sim y^{-3/2}\,F\!\left(y\,|\Delta|^{2}\right)}\;.
\end{equation}

\subsubsection{Total time-delay $\mathcal{A}[H(X)]$}\label{subsubsec:delay}

From Eqn.~ \ref{eqn:elem}, the delay is the sum of the elementary jams. Since Elementary jams are geometric features ({\it i.e.} the FCT) of $H(X)$, the total delay is a functional of $H(X)$, and we use $\mathcal{A}[H(X)]$ to denote the total-delay for a given initial condition associated with $H(X)$. 

\begin{figure}
    \centering
    \begin{tabular}{c}
        \includegraphics[width=0.75\linewidth]{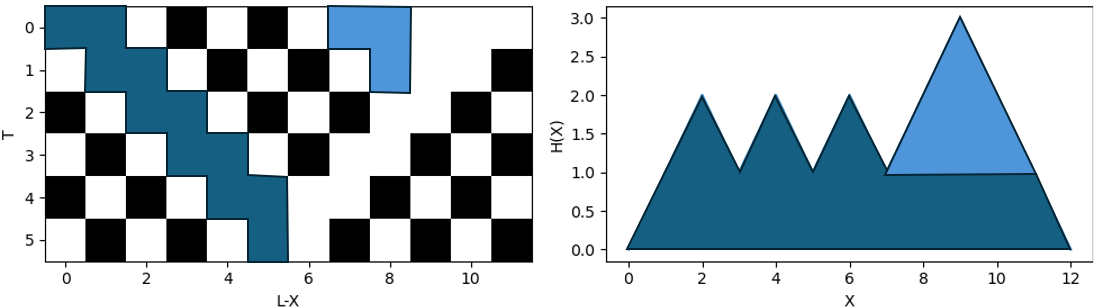} \\[-0.3em]
        \textbf{(a)} \\[0.8em]
        \includegraphics[width=0.75\linewidth]{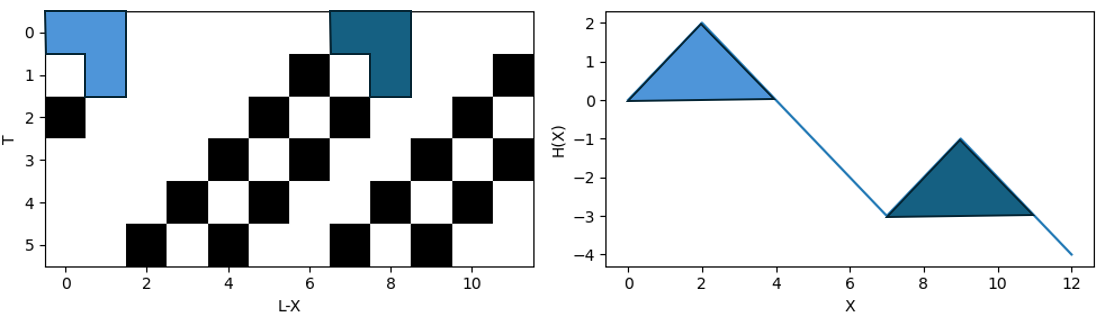} \\[-0.3em]
        \textbf{(b)} \\[0.8em]
        \includegraphics[width=0.75\linewidth]{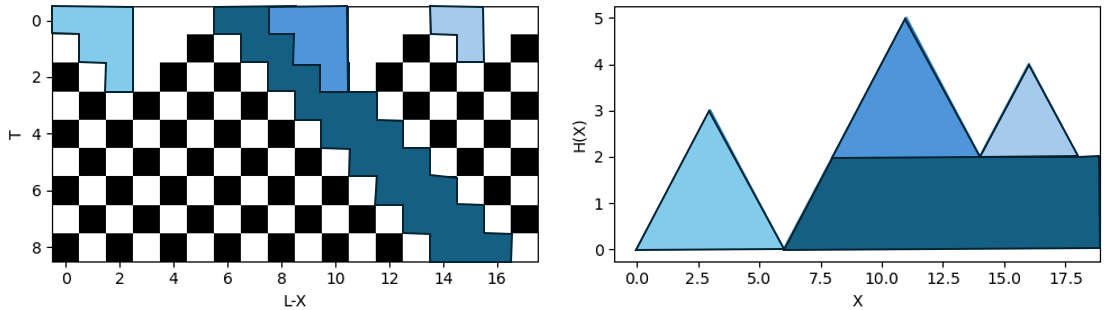} \\[-0.3em]
        \textbf{(c)}
    \end{tabular}
    \caption{The total delay $\mathcal{A}[H(X)]$ (sum of the colored cluster areas in the space-time plots) and its relationship to the area elements of $H(X)$ (via matching colors) : a) For $\rho_c = 0.5$, $\mathcal{A}[H(X)]$ is given by half the area under the curve $H(X)$. b) For $\rho<0.5$, $\mathcal{A}[H(X)]$ is given by half the area under the ``triangular tents" of $H(X)$. c) For $\rho > 0.5$, the area of the terminating jam clusters is given by the ``triangular tents" portions, and the non-terminating clusters contribute to infinitely extending horizontal area elements as shown. Note:  the space-time plots are presented with the X-axis flipped.}
    \label{fig:ax_hx}
\end{figure}

At the critical value, $\rho_c = 0.5$, $H(X)$ is a non-negative curve that starts and ends at $H = 0$. Thus at $\rho_c$, the FCT may be interpreted as horizontal sections of the region under the curve $H(X)$, for example see Fig.~\ref{fig:ax_hx} (a), and the total delay $\mathcal{A}[H(X)]$ is given by half the area under the curve $H(X)$ \cite{laval2023self}. In the continuum limit we may write,

\begin{align*}\label{eqn:delay_half}
    \mathcal{A}[h(x),\rho = \rho_c] = \frac{1}{2}\int_{0}^{\infty}h(x)dx
\end{align*}
Thus we get the average at $\rho_c$  ,

\begin{align}
    \langle \mathcal{A}(\rho=\rho_c)\rangle = \langle \mathcal{A}(\Delta=0)\rangle = \int \mathcal{D}h\{p[h(x)]\:( \frac{1}{2}\int_{0}^{\infty}h(x)dx)\}
\end{align}

For values of $\rho\neq\rho_c$, while the delay $\mathcal{A}[H(X)]$ is not related to the conventional area under the curve, it is still given by half the area of certain portions under the curve in the 2D H(X)-X plane (see Figs.~\ref{fig:ax_hx} b and c ). These portions are regions under the \textit{tent-like} parts of the curve where the first-crossing times are present. Thus in the continuum limit we write it simply as the function $\mathcal{A}[h(x)]$, i.e. 

\begin{align}\label{eqn:delay_general}
    \langle \mathcal{A}(\Delta)\rangle = \int \mathcal{D}h\{p[h(x)]\:\mathcal{A}[h(x)]\}
\end{align}

Strictly speaking, for a finite system of size $L$, the geometric area under $H(X)$ slightly under estimates the delay of the corresponding cluster. However as the system size increases, we expect the relative difference between the two vanishes in the limit $L\rightarrow\infty$. In Appendix~\ref{check_delay} we provide a simple example of this, and numerically probe the relative difference between the total delay and half the area under $H(X)$ for different $L$.

\subsubsection{Relaxation time $T_R$}\label{subsubsec:t_relax}

From Sec.~\ref{subsec:setup_phenom}, for densities $\rho \le \rho_c$, the relaxation time $T_R$ is given by the maximum jam cluster lifetime i.e.

\begin{align}\label{eqn:relaxn_time2}
    T_R = \max_i \:\theta_{i}, \qquad \rho <\rho_c
\end{align}

Suppose there are $N$ jam clusters in the system. Then the probability that the maximum lifetime $\theta_i$ is at most equal to $T_R$ is given by the joint probability that all of the $N$ elementary jams are less than $T_R$,

\begin{equation}
    P(\max\; \theta \leq T_R) = P(\cap_{i=1}^{N}\:\theta_i\leq T_R )
\end{equation}
In the limit $L\rightarrow\infty$ we can take these $N\rightarrow\infty$ cluster-lifetimes as being chosen independently, so that the cumulative probability of observing a relaxation time $T_R$ becomes the product of $N$ cumulative lifetime distributions:

\begin{equation}
    \lim_{L\rightarrow\infty}P(\max \theta \leq T_R) = \lim_{N\rightarrow\infty}\Pi_{i=1}^NP(\theta_i\leq T_R ) = \lim_{N\rightarrow\infty}(P(\theta\leq T_R ))^N,
\end{equation}
The cumulative probability of the lifetime is given by,

\begin{equation}
    P(\theta\leq T_R) = 1- \bar{S}(T_R) = 1- \int_{0}^{\infty}p(h',x = T_R)dh'
\end{equation}
where $\bar{S}(x)$ is the survival probability of the stochastic process Eqn.~\ref{eqn:fp_x} not reaching $h=0$ by time $x$. Thus,

\begin{equation}
   \lim_{L\rightarrow\infty} P(\max \theta \leq T_R) = \lim_{N\rightarrow\infty}(P(\theta\leq T_R ))^N = \lim_{N\rightarrow\infty} (1- \int_{0}^{\infty}p(h', T_R)dh')^N,
\end{equation}

and the probability that the maximum $\theta = T_R$ is

\begin{equation}\label{eqn:p_Tr}
    \bar{p}(T_R) = \lim_{N\rightarrow\infty} \frac{d}{dT_R}(( 1- \int_{0}^{\infty}p(h',T_R)dh')^N)
\end{equation}
where $\bar{p}(T_R)$ is the probability density function of $T_R$ at density $\rho$.   Thus,
\begin{align}\label{eqn:tr_avg_eqn}
    \langle T_R\rangle = \int_{0}^{\infty}T_R\: \bar{p}(T_R) dT_R
\end{align}

Finally, note that the number of clusters $N$ is also a function of $\Delta$ and $L$;  however, in Appendix~\ref{check_tr} we provide numerical evidence justifying the use of a constant $N$ in the above calculation.



\section{Scaling analysis and Critical exponents}\label{sec:scalexp}

\subsection{Scaling forms of $p(h,x)$ and  $p[h(x)]$}

The solution $p(h,x,\Delta)$ to the Fokker-Planck Eqn.~\ref{eqn:fp_x} is given by Eqn.~\ref{eqn:fp_soln}. Unfortunately, substituting it and solving for the macroscopic observables in Eqns.~\ref{eqn:delay_general} and \ref{eqn:p_Tr} is not straighforward. On the other hand, we observe that $p(h,x,\Delta)$ is a homogeneous function with the following scaling property for any $s>0$:

\begin{align}\label{eqn:p_scaling}
   \boxed{ p(h,x,\Delta) = s^{1/2}p(s^{1/2}h,sx,\frac{\Delta}{s^{1/2}}) }
\end{align}
Similarly $p[h(x),\Delta]$ satisfies the scaling form (details in Appendix~\ref{app:scalingDerivation} ):

\begin{align}\label{ph_scaling}
    p[h(x), \Delta] = \frac{1}{\mathcal{J}(s)}p[s^{1/2}h(sx), \frac{\Delta}{s^{1/2}}]
\end{align}
where $\mathcal{J}(s)$ is the Jacobian for the transformation $h\longrightarrow h' = \sqrt{s}h$ with $\mathcal{D}h = \mathcal{J}(s)\mathcal{D}h'$.


\subsection{Critical exponents via scaling forms}

From Sec.~\ref{subsubsec:t_relax} we know that $\bar{p}(T_R,\Delta)$ is given by the eqn~\ref{eqn:p_Tr}, and using eqn~\ref{eqn:p_scaling} we get

\begin{align*}
    \bar{p}(T_R,\Delta) = \lim_{N\rightarrow\infty} s\frac{d}{d(sT_R)}(( 1- \int_{0}^{\infty}\sqrt{s}p(\sqrt{s}\;h,sT_R,\frac{\Delta}{\sqrt{s}})\frac{d(\sqrt{s}h)}{\sqrt{s}})^N)
\end{align*}

The change of variables

\begin{align*}
    T_R' & = sT_R\\
    h' & = h\sqrt{s}
\end{align*}

gives 

\begin{align*}
    \bar{p}(T_R,\Delta) = \lim_{N\rightarrow\infty} s\frac{d}{dT_R'}(( 1- \int_{0}^{\infty}p(h',T_R',\frac{\Delta}{\sqrt{s}})dh')^N) 
\end{align*}

This may be rewritten as a scaling property for $\bar{p}(T_R,\Delta)$:

\begin{align}\label{eqn:tr_scale_awesome}
    \boxed{\bar{p}(T_R,\Delta) = s \bar{p}(sT_R,\frac{\Delta}{\sqrt{s}})}
\end{align}

\begin{figure}
    \centering
    \includegraphics[width=0.7\linewidth]{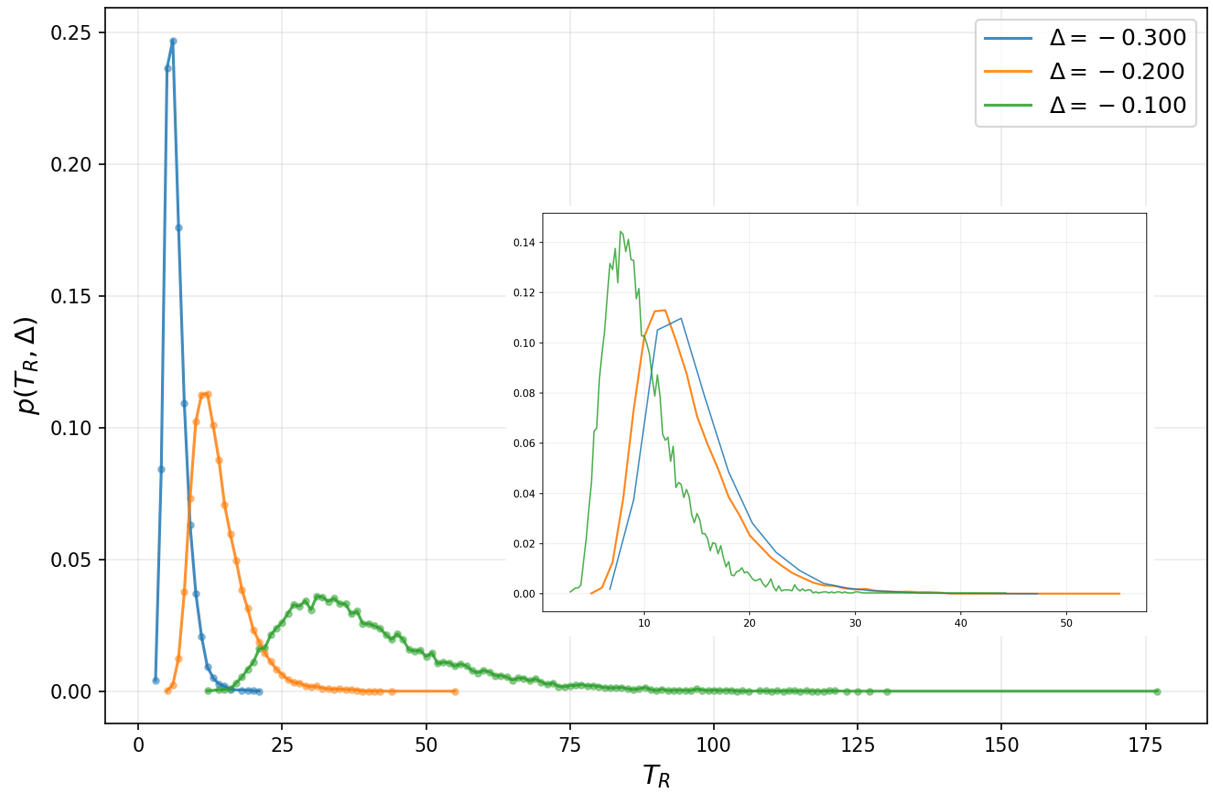}
    \caption{Relaxation-time distribution $p(T_R,\Delta)$ for a system of size $L = 1000$ with a sample of $10000$ simulations each for $\Delta = \{-0.1,-0.2,-0.3\}$. (inset) Data collapse using the scaling form $\bar{p}(T_R,\Delta) = s\,\bar{p}(sT_R,\Delta/\sqrt{s})$. The reference curve is set at $\Delta = -0.200$; data at $\Delta = -0.300$ and $\Delta = -0.100$ are rescaled with $s = (\Delta / \Delta_{\rm ref})^{2}$, giving $s = 2.25$ and $s = 0.25$, respectively. The curves approximately collapse onto a single scaling function, with some deviation possibly due to finite size of the simulation. }
    \label{fig:tr_awesome_scale}
\end{figure}

A computational test of  Eqn.~\ref{eqn:tr_scale_awesome} is shown in Fig.~\ref{fig:tr_awesome_scale}.

Next, we use eqn~\ref{eqn:tr_scale_awesome} to write

\begin{align}\label{eqn:tr_scaling_form}
    \langle T_R(\Delta) \rangle &= \int_{0}^{\infty} T_R\; \bar{p}(T_R,\Delta)dT_R \\
    &= \int_{0}^{\infty} sT_R\; \bar{p}(sT_R,\frac{\Delta}{\sqrt{s}})dT_R \\
    &= \frac{1}{s}\int_{0}^{\infty} T_R'\; \bar{p}(T_R',\frac{\Delta}{\sqrt{s}})dT_R'\\
    &= \frac{1}{s}\langle T_R (\frac{\Delta}{\sqrt{s}})\rangle 
\end{align}
Since $s$ is an arbitrary positive constant, it follows that 

\begin{align}
\label{Tr_critical}
\boxed{
    T_R(\Delta) \sim \Delta^{-2}
}    
\end{align}

Similarly, for the total delay $\langle A\rangle$ we write
\begin{align}
    \langle A(\Delta)\rangle
    &= \int \mathcal{D}h \;\mathcal{A}[h]\; p[h,\Delta].
\end{align}

Since this quantity is proportional to subsections of the 2D area under $h(x)$ (Sec.~\ref{subsubsec:delay}), under the transformation 
\[
    x'=sx, \qquad h'(x')=\sqrt{s}h(x),
\]
the delay functional scales similarly to any 2D area element, $dA = dh\;dx$ , {\it i.e.} 
\[
    \mathcal{A}[h]  =  s^{-3/2}\,\mathcal{A}[h'],
\]

Using the scaling form for 
$p[h,\Delta]$, we obtain
\begin{align}\label{eqn:scaling_A}
    \langle A(\Delta)\rangle
    &= \int \mathcal{D}h'\,\mathcal{J}(s)\,s^{-3/2}\mathcal{A}[h']
       \; \frac{1}{\mathcal{J}(s)} p\!\left[h',\frac{\Delta}{\sqrt{s}}\right] \\
    &= s^{-3/2}\int\mathcal{D}h'\,\mathcal{A}[h']\;
       p\!\left[h',\frac{\Delta}{\sqrt{s}}\right] \\
    &= s^{-3/2}\,\langle A\!\left(\frac{\Delta}{\sqrt{s}}\right)\rangle.
\end{align}

Again, since $s$ is an arbitrary positive constant, we have 

\begin{align}
\label{eqn:A_critical}
\boxed{
    \langle A(\Delta)\rangle \sim \Delta^{-3}
}    
\end{align}

(Since $0 < \rho < 1$, the value of $\Delta$ in the argument is restricted to have magnitude between $0$ and $0.5$ .)

Continuing, substitute  $s = \frac{1}{L}$ into  Eqn.~\ref{eqn:scaling_A} , which yields the finite-size dependence of $A$:

\begin{align}\label{A_fss}
    \langle A(\Delta)\rangle
    = L^{3/2}\,\langle A\!\left(L^{1/2}\Delta\right)\rangle.
\end{align}

Thus, at the critical point $\Delta = 0$ , $\langle A(0)\rangle \sim L^{3/2}$, which is consistent with \cite{majumdar2005airy,laval2023self}. Defining the normalized order parameter $\phi = 2A/L^{2}$ (recall that the maximum area of the space--time plot is $L^{2}/2$) and substituting in Eqn.~\ref{A_fss} yields the finite-size scaling form for $\phi$ 
\begin{equation}
    \phi = L^{-1/2}g(L^{1/2}(\Delta))
    \label{phi_fss}
\end{equation}
which is precisely the relation deduced from earlier numerical simulations \cite{jha2025simple}.
Substituting $L = 1/s$ back into Eqn. \ref{phi_fss} above and using $s \sim \Delta^{2}$, we obtain the critical exponent for the order parameter,
\[
\boxed{\phi \sim \Delta^{1}}.
\]



\section{Summary and discussion}\label{sec:summart_disc}

We developed an analytic statistical description of the transient dynamics of the deterministic traffic model ECA184.  Along the way, we introduced the height function, and obtained its time independent probability distribution over initial lattice configurations,  reminiscent of a ``micro-cannonical" distribution in equilibrium statistical mechanics. 

In fact, our analysis admits an effective equilibrium thermodynamic description in space--time. Nagel and Paczuski~\cite{nagel1995emergent} proposed a phenomenological random walk model for the time evolution of a traffic jam of size $n(t)$, with probability density $P(n,t)$ that satisfies
\begin{align}
\label{NP_pde}
\frac{\partial P}{\partial t}
= (r_{\mathrm{out}}-r_{\mathrm{in}})\frac{\partial P}{\partial n}
+ \frac{r_{\mathrm{out}}+r_{\mathrm{in}}}{2}\frac{\partial^2 P}{\partial n^2},
\end{align}
where $r_{\mathrm{in}}$ and $r_{\mathrm{out}}$ are taken as density–dependent rates at which vehicles enter and leave the jam.

Under the identification
\[
x \rightarrow t, \qquad h \rightarrow n,\qquad p\rightarrow P
\]
and the parameter mapping
\[
-(2\rho-1)v_0 \rightarrow (r_{\mathrm{out}}-r_{\mathrm{in}}),
\qquad
D_0 \rightarrow \frac{r_{\mathrm{out}}+r_{\mathrm{in}}}{2},
\]
Eqn.\ref{NP_pde} the above equation is formally identical to our Fokker--Planck equation
\begin{align}\label{eqn:fp}
\frac{\partial p}{\partial x}
= -(2\rho-1)v_0\frac{\partial p}{\partial h}
+ D_0\frac{\partial^2 p}{\partial h^2}.
\end{align}

This formal mapping is physically meaningful. In the phenomenological description, the first return to $n=0$ determines the jam lifetime distribution. In our approach, the same first-return statistics arise automatically and naturally. Similarly, the total delay caused by a jam is given by $\int n(t)\,dt$ up to the first return time, corresponding to the area under the trajectory, which also appears naturally in our formulation.

Identifying
\[
r_{\mathrm{in}} = \rho v_0, \qquad
r_{\mathrm{out}} = \rho(1-v_0),
\]
we obtain
\[
r_{\mathrm{out}}-r_{\mathrm{in}} = -(2\rho-1)v_0,
\qquad
D_0 = \frac{r_{\mathrm{in}}+r_{\mathrm{out}}}{2} = \frac{v_0}{2} .
\]
Thus, the parameter $D_0$ admits a physical interpretation as a symmetrized flow rate.

Despite the formal similarity, there are significant differences. First, the approach of Nagel and Paczuski is phenomenological, whereas our results are exact consequences of ECA184's deterministic dynamics.  Second, the assumption that $r_{\mathrm{in}}$ and $r_{\mathrm{out}}$ depend only on the control parameter $\rho$ is a mean–field approximation.  Their phenomenological equation applies to a single isolated jam, while our method applies to the entire lattice consisting of interacting jams and voids. On the other hand, Nagel and Paczuski consider a more general case of accelerating vehicles with $v_0>1$, which we do not consider in our analysis.


We were able to determine critical exponents for the microscopic observables. Then, exploiting scaling arguments, we derived critical exponents for the macroscopic observables, as well as information about microscopic jam statistics. Next, we discuss why we used different methods to derive the critical exponents for macroscopic and microscopic observables.  To determine the latter, we relied on an explicit solution to the Fokker-Planck equation; for the former, we only needed the scaling forms.  Now, we were able to use scaling forms to solve for the macroscopic averages by substituting $s \sim \Delta^2$ in Eqns.~\ref{eqn:tr_scaling_form} and ~\ref{eqn:scaling_A}, because $\Delta$ is dimensionless:   $\rho$ is introduced as a probability in Eqn.~\ref{eqn:markov_thermo}, and is thus dimensionless in this problem. Any substitution to $s$ in a scaling form which is dimensionless -- and physically meaningful, {\it i.e.} say $\Delta' = \frac{\Delta}{\sqrt{s}}\epsilon(-\frac{1}{2},\frac{1}{2})$) -- can be applied to scaling forms. Thus, we obtain the correct critical exponents given by relations Eqns.~\ref{eqn:A_critical} and ~\ref{Tr_critical}. Similarly, we obtain valid finite size scaling results in Eqn.~\ref{phi_fss} because $L\sim\Delta^{-1}$ is also dimensionless.

Conversely, the microscopic observables $\theta$ and $m$ are not dimensionless, so we do not get the correct exponents for the microscopic distribution $p'(y,\Delta)$, $y =\{m,\theta\}$ using scaling forms.  

To see what goes wrong, note first that we can obtain the scaling form for $p'(\cdot)$, using the relation between first return time and the survival function:

\begin{align}
    p'(\theta,\Delta) 
    &= -\int_{0}^{\infty}\frac{\partial}{\partial \theta}p(h,2\theta,\Delta)dh \\
    &= -\int_{0}^{\infty}s\frac{\partial}{\partial (s\theta)}\;p(\sqrt{s}h,2(s\theta),\frac{\Delta}{\sqrt{s}})d(\sqrt{s}h)\\
    &= sp'(s\theta,\frac{\Delta}{\sqrt{s}})
\end{align}

But substituting $s = \theta^{-1}$ at $\Delta=0$ does not yield the $-\frac{3}{2}$ exponent we obtained in Sec.~\ref{microscopic_distr_expo}.  This is because $\theta$ is not dimensionless:  it has dimensions of space ``$x$" (the other dimensional variable is the height $h$).


Finally, we return to the question of the  multiplicity of the mapping between the initial conditions and the height function $h(x)$.   Recall that we set $w[h(x)] = 1$ in the limit $L\rightarrow\infty$ in Eqn.~\ref{eqn:weighted}, which we justify through the concept of \textit{typicality} \cite{cover1999elements}. The initial conditions are elements of the Bernoulli ensemble, {\it i.e.} binary sequences of length $L$. 
 In the limit $L\rightarrow\infty$, the maximum weight of the sampled initial conditions lies in the \textit{typical set}. In particular, the different sequences in the typical set are all related to each other via a shift operation. Thus, we expect that with weight $w[h_0(x)] = 1$, the initial conditions map to the same $h_0$. Of course, this $h_0$ is special in that it captures the entire range of statistics of $h(x)$ in a single curve.

\section*{Declarations}

\noindent\textbf{Funding: }This research was funded in part by NSF Award No. 2311159.
\newline
\noindent\textbf{Conflict of interest: } The authors declare that there are no conflict of interest.
\newline
\noindent\textbf{Data and Code availability: }The data supporting the findings in this article are openly
available on GitHub \cite{Jha2025ECA184Critical}
\newline


\noindent\rule{\linewidth}{0.4pt}

\appendix

\renewcommand{\appendixname}{\MakeUppercase{Appendix}} 
\renewcommand{\thesection}{\Alph{section}} 


\section{Numerical Validation for expressions of macroscopic observables via $H(X)$}\label{app:appA}

\subsection{Total delay and area under $h(x)$}\label{check_delay}

In Sec.~\ref{subsubsec:delay} we stated that in Eqn.~\ref{eqn:delay_half}, at $\rho = 0.5$, the total delay $\mathcal{A}$ is related to $h(x)$ as half the area under the curve.  Here, we provide examples and numerical checks for this relation. We note that it is true only in the limit of large clusters (and large system sizes).

\subsubsection{ Example via a simple cluster shape}

We examine how the cluster delay and value of $\frac{1}{2}\int h(x)dx$ differs for a simple triangular cluster. 

\begin{figure}
    \centering
    \includegraphics[width=0.8\linewidth]{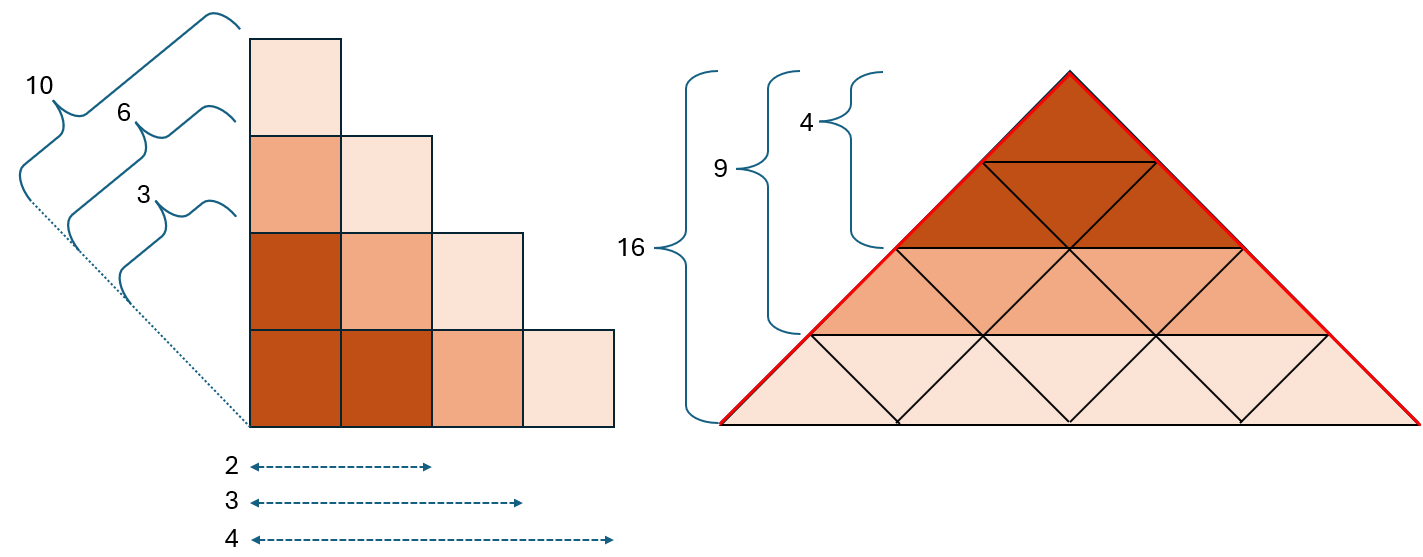}
    \caption{(Left) Triangular jam clusters of cluster size 3, 6 and 10 arising from initial jams of run length $n =$ 2, 3 and 4 respectively. We obtain a jam delay of $\frac{n^2+n}{2}$. (Right) Corresponding height functions $H(X)$ mapped via color components. We find $n^2$ triangles each of area 1 i.e. area  $A[H(X)] = n^2$.}
    \label{fig:hx_area_example}
\end{figure}

Fig~\ref{fig:hx_area_example} shows a cluster and its associated triangular height function $H(X)$. We can use this cluster shape as well as its height function to test  the relation between the delay and $H(X)$ area for any such cluster formed from an initial jam of length $n$. 

The area under $h(x)$ is equal to the number of small triangles each of area 1.  Let $n$ denote the initial jam length.  For $n = 2$:  the \text{delay} = 3, the number of triangles = $4$, and the area under $H(X)= 4$. For $n=3$: the \text{delay} = $6$, the number of triangles = 9, and the area under $H(X)$ = 9.  For general $n$, the \text{delay} = $\tfrac{1}{2} \left(n^2 + n\right)$, and the number of triangles is equal to the area under $H(X)$ = $n^2$.  

Thus at large $n$ for a triangular shaped jam cluster,

\begin{align}
    \text{delay} \approx \frac{1}{2}\text{area under}\,{H(X)} \sim \tfrac{1}{2}n^2.
\end{align}

\subsubsection{Numerical check via sampling }

\begin{figure}
    \centering
    \includegraphics[width=0.5\linewidth]{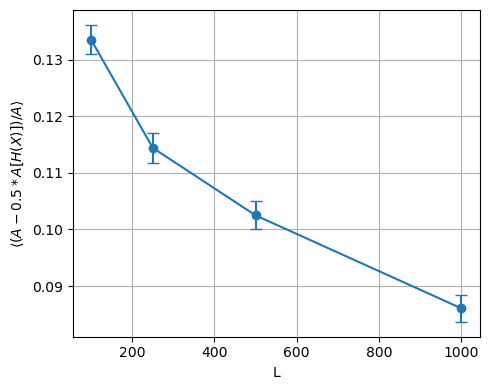}
    \caption{Mean relative difference $\left(A - \tfrac{1}{2}A_H\right)/A$ as a function of system size $L$ at $\rho = 0.5$, averaged over $M = 1000$ initial conditions. Error bars denote the standard error.}
    \label{fig:delay_num}
\end{figure}

At the critical density $\rho = 0.5$, we numerically compare the total delay with the area under the height function $H(X)$ for finite systems of size $L = 100, 250, 500, 1000$. 
For each system size, we sample $M = 1000$ independent initial conditions with exactly half the sites occupied. 
For each realization, we construct the height function $H(X)$, compute its discrete area $A_H$ and independently use the simulation to directly evaluate the total delay $A$. 
We then quantify the discrepancy through the relative difference
\[
\delta_{\text{rel}} = \frac{A - \frac{1}{2}A_H}{A},
\]
where $A$ denotes the total delay as measured from the simulation.  Finally, for each $L$ , we average $\delta_\text{rel}$ over the $M$ realizations.  (Fig.~\ref{fig:delay_num}) shows the result.  The mean relative difference decays systematically with $L$, providing numerical evidence that $\tfrac{1}{2}A_H$ converges to the total delay $A$ in the large-$L$ limit.

\subsection{$\bar{p}(T_R,\Delta)$ and $N(\Delta,L)$}\label{check_tr}

In Sec.~\ref{subsubsec:t_relax} we derived the relaxation-time distribution
\begin{align}
\bar{p}(T_R,\Delta)
= \lim_{N\to\infty}\frac{d}{dT_R}\Big(1-\int_{0}^{\infty} p(h,T_R,\Delta)\,dh\Big)^{N}.
\label{eq:pbar_def}
\end{align}
where $N$ is the number of distinct clusters. In this derivation $N$ was taken to be large and effectively independent of both $\Delta$ and $L$.

More generally, however, we write $N= N(\Delta,L)$ .  We now show that this dependence does not affect the validity of~\eqref{eq:pbar_def}.

\begin{figure}[t]
    \centering
    \begin{subfigure}{0.48\linewidth}
        \centering
        \includegraphics[width=\linewidth]{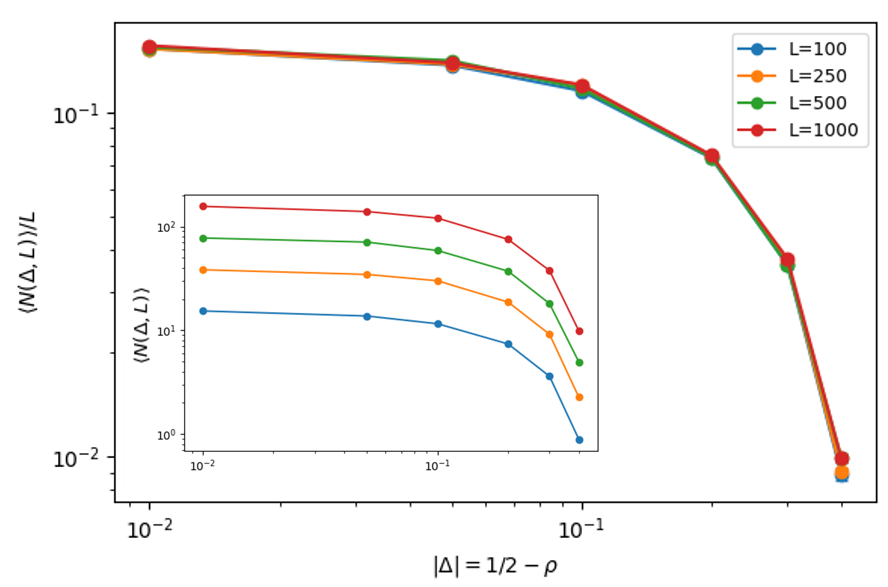}
        \caption{}
        \label{fig:TR_numeric}
    \end{subfigure}
    \hfill
    \begin{subfigure}{0.45\linewidth}
        \centering
        \includegraphics[width=\linewidth]{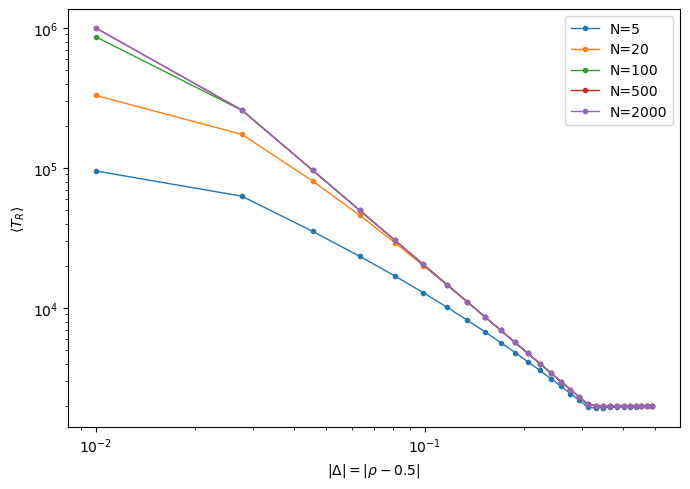}
        \caption{}
        \label{fig:TR_convergence}
    \end{subfigure}
    \caption{(a) Collapse of the mean number of clusters when plotting $N(\Delta,L)/L$ versus $|\Delta|$ for different system sizes $L$, averaged over $M=1000$ initial conditions, demonstrating the factorized form $N(\Delta,L)\simeq L\,N_1(\Delta)$. The inset shows the original, non-collapsed data for $N(\Delta,L)$ as a function of $|\Delta|$.(b) Convergence of $\langle T_R\rangle$ with increasing number of clusters $N$ for fixed $|\Delta|$, showing that $\bar{p}(T_R,\Delta)$ becomes independent of $N$ already for moderate $N$. We also note that the system approaches a power law with slope $-2$ as expected from calculations.}
    \label{fig:TR_checks}
\end{figure}

First, for $\Delta \neq 0$ we find numerically -- see Fig.~\ref{fig:TR_checks}(a) -- that the mean number of clusters factors as
\begin{align}
N(\Delta,L) \simeq N_1(\Delta)\,N_2(L),
\qquad N_2(L) \sim L ,
\label{eq:N_factor}
\end{align}
so that $N(\Delta,L)$ grows linearly with system size while $N_1(\Delta)$ remains finite for fixed $\Delta$.   In particular, in the limit $L\to\infty$,  $N(\Delta,L) = \infty$ for any $\Delta \neq 0$.

Second, we verify directly that $\langle T_R\rangle$ converges rapidly as a function of $N$. Using Eqn.~\ref{eqn:tr_avg_eqn} and substituting the solution for $p(\cdot)$ in Eqn.~\ref{eqn:fp_soln} we numerically obtain $\langle T_R\rangle$ as shown in Fig.~\ref{fig:TR_checks}(b). The numerical estimates of $\langle T_R\rangle$ become essentially independent of $N$ already for moderate values of $N$. From Eqn.~\ref{eq:N_factor} the large $L$ limit ensures that $N$ is large enough.

We therefore conclude that the dependence $N=N(\Delta,L)$ does not modify the form of $\bar{p}(T_R,\Delta)$ obtained in Sec.~\ref{subsubsec:t_relax}.

\section{Derivation of Scaling form of $p[h(x)]$}\label{app:scalingDerivation}

Here we derive the scaling form of $p[h(x),\Delta] = \frac{1}{Z(\Delta)}\exp\{-S[h(x),\Delta]\}$.

First, apply the transformation  $x = x'/s$ and $h(x) = h'(x')/\sqrt{s}$ to get 

\begin{align*}
    S[h(x),\Delta]  & = \int_{x_0}^{x}dx [K_2 (\partial_{x}h)^2+K_0] \\
    & = \int_{sx_0}^{sx}\frac{dx'}{s} [K_2 (\sqrt{s}\partial_{x'}h')^2+K_0] \\
    & = \int_{x_0'}^{x'}dx' [K_2 (\partial_{x'}h')^2+\frac{K_0}{s}]\\
    & = S[\sqrt{s}h(sx), \frac{\Delta}{\sqrt{s}}]
\end{align*}

The transformation $\mathcal{D}h \rightarrow \mathcal{D}h'$ is obtained from
$\mathcal{D}h = \prod_i dh(\tau_i)$ together with the scaling
$h'(x') = h(x)/\sqrt{s}$.  Thus,
\begin{align*}
    dh(x_i) &= \sqrt{s}\, dh'(x_i'),\\[0.3em]
    \mathcal{D}h
    &= \Bigl(\prod_i \sqrt{s}\Bigr)\mathcal{D}h'
     = \mathcal{J}(s)\,\mathcal{D}h',
\end{align*}
where $\mathcal{J}(s)$ is the Jacobian of the transformation.

It follows that 

\begin{align*}
    Z(\Delta) &= \int \mathcal{D}h(x) \exp\{-S[h(x),\Delta]\}\\
    & = \mathcal{J}(s)\int\mathcal{D}h'\exp\{-S[h'(x'), \frac{\Delta}{\sqrt{s}}]\}\\
    & = \mathcal{J}(s) Z(\frac{\Delta}{\sqrt{s}})
\end{align*}

Combine these results to obtain the scaling for $p[h(x),\Delta]$ :

\begin{align}\label{ph_scaling}
    p[h(x), \Delta] & = \frac{1}{Z(\Delta)}\exp\{-S[h(x),\Delta]\}\\
    & = \frac{1}{\mathcal{J}(s)Z(\frac{\Delta}{\sqrt{s}})}\exp\{-S[\sqrt{s}h(sx), \frac{\Delta}{\sqrt{s}}]\} \\
    & = \frac{1}{\mathcal{J}(s)}p[s^{1/2}h(sx), \frac{\Delta}{s^{1/2}}]
\end{align}
%




\renewcommand{\bibsection}{\begin{center}\rule{0.5\textwidth}{0.4pt}\end{center}}
\setcounter{page}{1}
\renewcommand\refname{References Cited}
\bibliography{references}
\bibliographystyle{unsrtnat} 

\end{document}